\documentclass[aps,preprint,showpacs,floatfix]{revtex4}
\usepackage{graphicx}% Include figure files
\usepackage[latin1]{inputenc}
\usepackage[T1]{fontenc}
\usepackage[english]{babel}
\usepackage{epsfig}
\usepackage{amssymb}
\usepackage[centertags]{amsmath}
\usepackage{amsthm}
\usepackage{amsfonts}
\usepackage{makeidx}
\usepackage{graphics}
\usepackage{graphicx}
\usepackage{psfrag}
\usepackage{subfigure}
\usepackage{dsfont}
\usepackage{eucal}
\usepackage{color}
\usepackage{bm}

\def\bold#1{\setbox0=\hbox{$#1$}%
     \kern-.025em\copy0\kern-\wd0
     \kern.05em\%\baselineskip=18ptemptcopy0\kern-\wd0
     \kern-.025em\raise.0433em\box0 }
\def\slash#1{\setbox0=\hbox{$#1$}#1\hskip-\wd0\dimen0=5pt\advance
         to\wd0{\hss\sl/\/\hss}}
\def \de {\partial}
\def \a {\alpha}
\def \b {\beta}
\def \g {\gamma}
\def \G {\Gamma}
\def \d {\delta}

\def \ve {\varepsilon}

\def \l {\lambda}

\def \m {\mu}
\def \n {\nu}

\def \s {\sigma}

\def \zh {\hat z}
\def \be {\begin{equation}}
\def \ee {\end{equation}}
\def \bea {\begin{eqnarray}}
\def \eea {\end{eqnarray}}
\def \non {\nonumber}
\def \noi {\noindent}
\def \ra {\rightarrow}

\def \fr {\displaystyle\frac}

\def\laq{~\raise 0.4ex\hbox{$<$}\kern -0.8em\lower 0.62
ex\hbox{$\sim$}~}
\def\gaq{~\raise 0.4ex\hbox{$>$}\kern -0.7em\lower 0.62
ex\hbox{$\sim$}~}

\def \wt {\widetilde}

\def \A {{\CMcal A}}

\begin{document}
%\begin{titlepage}
\addtolength{\jot}{10pt} \addtolength{\jot}{10pt}

\preprint{\vbox{\hbox{BARI-TH/08-593 \hfill}}}

\vspace*{1cm}

\title{\bf Light scalar mesons in the soft-wall model of AdS/QCD\\}

\author{P. Colangelo$^a$, F. De~Fazio$^a$, F. Giannuzzi$^{a,b}$, F. Jugeau$^a$ and S. Nicotri$^{a,b}$}

\affiliation{ $^a$ Istituto Nazionale di Fisica Nucleare, Sezione di Bari, Italy\\
$^b$ Dipartimento di Fisica,  Universit\`a  di Bari, Italy\\}

\begin{abstract}
We study light scalar mesons in the AdS/QCD soft-wall model with
a background dilaton field. The masses and decay constants are
compatible with experiment and QCD determinations if $a_0(980)$
and $f_0(980)$ are identified as the lightest scalar mesons;
moreover, the states are organized in linear Regge trajectories
with  the same slope of vector mesons. Comparing the two-point
correlation function of scalar operators in AdS and QCD,
information about the condensates can be derived. Strong
couplings of scalar states to pairs of light pseudoscalar mesons
turn out to be small, at odds with experiment and QCD estimates:
this discrepancy is related to the description of chiral
symmetry breaking  in this model.
\end{abstract}

\vspace*{2cm}


\maketitle
%\end{titlepage}

\newpage
%\setcounter{page}{1}
%\tableofcontents

\section{Introduction}
The idea of extending the AdS/CFT correspondence conjecture
\cite{Maldacena:1997re} to QCD-like theories \cite{Witten2:1998}
has provided new hints on the possibility of describing strong
interaction processes by string-inspired approaches. Two main
ways have been followed to achieve such a result. The first, the
so-called top-down approach, consists in starting from a
string/M-theory living on AdS$_{d+1}\times{\CMcal C}$ (${\CMcal
C}$ being a compact manifold) and attempting a derivation of a
low-energy QCD-like theory on the flat boundary ${\CMcal M}_d$
of the AdS space through appropriate compactifications of the
extra dimensions \cite{Witten2:1998,top-down}. In the second
one, the so-called bottom-up approach, one starts from $4d$ QCD
and attempts to construct its higher dimensional dual theory
(not necessarily a string one), assuming its existence
\cite{polchinsky}, with phenomenological properties as
guidelines.

An important aspect of both these approaches is the necessity of
a mechanism to break conformal invariance, since QCD is not a
conformal theory \cite{confqcd},  and to account for phenomena
such as confinement. The way to do this usually consists in
incorporating in the dual theory a mass scale related to the QCD
scale $\Lambda_{QCD}$. For example, in the bottom-up approach,
one possibility is to use a five dimensional ``AdS-slice''
 letting the fifth (holographic) coordinate
$z$ vary in a range up to $z_{max}$ of ${\CMcal
O}(\frac{1}{\Lambda_{QCD}})$ {\cite{polchinsky,son1,pomarol1}.
In this (so-called hard-wall) model, several QCD aspects have
been investigated, namely high-energy hadron scattering
amplitudes, spectra, form factors, strong couplings, light-front
wave functions, Wilson loop
\cite{polchinsky,son1,pomarol1,AdSQCD1,braga}.

\noi Another proposal to break conformal invariance consists in
introducing in the $5d$ AdS holographic space a background
dilaton field (the so-called soft-wall model) \cite{Andreev soft
wall,son2}. While in top-down approaches the dilaton profile
must be a solution of the supergravity equations of motion, in
this kind of approaches its functional form is chosen on the
basis of phenomenological information, namely imposing the Regge
behaviour for vector mesons; noticeably, the obtained dilaton
profile,   found using heuristic arguments, can be justified
constructing a suitable dynamical model \cite{Batell:2008zm}.
Also in this framework many  QCD properties have been
investigated, such as vector and tensor meson masses and form
factors, glueball masses, the static $\bar QQ$ potential and DIS
\cite{son2,radyushkin,lebed,jugeau1,andreev,Andreev soft
wall,braga1}. The results of the two bottom-up models differ in
many respects; the possibility of continuously interpolating
between them has also been considered \cite{lebed2}, trying to
recognize the essential features of the QCD dual.

Since light scalar mesons represent an important and debated
sector of QCD, it is interesting to consider them in the
holographic framework, and indeed some analyses have been
carried out  \cite{pomarol2,vega}. Here we study the scalar
sector in the soft-wall model, trying to identify which
properties can be described in the holographic approach. In
particular, we consider the mass spectrum, the decay constants
and the strong couplings of scalar mesons to pairs of light
pseudoscalars. The comparison of the results obtained in the AdS
framework with experiment and QCD calculations can shed light on
the features and drawbacks of this model.

\section{The model}

The  model we investigate is defined in the $5d$ space with
metric:
\begin{equation}
ds^2=g_{MN} dx^M dx^N=\frac{R^2}{z^2}\,\big(\eta_{\m\n}dx^{\m}dx^{\n}+dz^2\big) \label{metric}
\end{equation}
with $\eta_{\m\n}=\mbox{diag}(-1,+1,+1,+1)$; $R$ is the AdS
curvature radius and the coordinate $z$  runs in the range
$0\leq z < +\infty$ (or, considering a UV cutoff, from the
ultra-violet brane $z_{min}=\epsilon$ to $+\infty$).

In addition to the AdS metric, the model is characterized by a
background dilaton field:
\begin{equation}
\Phi(z)=(c z)^2 \label{dilaton}
\end{equation}
the form of which is chosen  to obtain  light vector mesons
with linear Regge trajectories \cite{son2};  $c$ is a
dimensionful parameter setting the scale of QCD quantities.

We consider the $5d$ action:
\begin{equation}
S_{eff}=-\frac{1}{k}\int
d^5x\sqrt{-g}\,e^{-\Phi(z)}\,\mbox{Tr}\Big\{|DX|^2+m_5^2X^2+\frac{1}{4g_5^2}\big(F_L^2+F_R^2\big)\Big\}
\label{action}
\end{equation}
where $g$ is the determinant of the metric tensor $g_{MN}$ in
\eqref{metric} and $\Phi$ the background dilaton field
\eqref{dilaton}. This action includes  fields which are dual to
QCD operators defined at the boundary $z=0$. There is a scalar
bulk field $X$, the mass of which is fixed by the AdS/CFT
relation: $m_5^2R^2=(\Delta-p)(\Delta+p-4)$,  $\Delta$ being the
dimension of the $p-$form QCD operator dual to $X$. This field,
written as
\begin{eqnarray}
X=(X_0+S)e^{2i\pi}\,\,\,,
%X^{ij}=(X_0^{ik}+S^{ik})\big(e^{2i\pi}\big)^{kj}
\end{eqnarray}
contains a background field $X_0(z)=\frac{v(z)}{2}$, the  scalar
field $S(x,z)$ and the  chiral field $\pi(x,z)$.  $X_0$ only
depends on $z$  and  is  dual to $\langle \bar q q\rangle$;
since it is different from zero, it represents the term
responsible for the breaking of  chiral symmetry. The scalar
bulk  field  $S$ includes  singlet $S_1(x,z)$ and  octet
$S_8^a(x,z)$ components,  gathered into the multiplet:
\begin{equation}
S=S^AT^A=S_1T^0+S_8^aT^a
\end{equation}
with $T^0={1}/\sqrt{2n_F}={1}/\sqrt{6}$ and $T^a$ the generators
of $SU(3)_F$, with  normalization
\begin{equation}
\mbox{Tr}\Big(T^AT^B\Big)=\frac{\d^{AB}}{2}
\end{equation}
($A=0,a$, and $a=1,\ldots 8$). $S^A$   is  dual  to the QCD
operator $\CMcal{O}^A_S(x)=\overline{q}(x)T^A q(x)$, so that
$\Delta=3$, $p=0$ and  $m_5^2R^2=-3$. The fact that the scalar
bulk field is tachyonic does not affect the stability of the
theory, since fields with slightly negative masses are allowed,
as discussed in \cite{Klebanov:1999tb}.

The action \eqref{action} also involves the  fields
$A^a_{L,R}(x,z)$ introduced to gauge the chiral symmetry in the
$5d$ space. They  are dual to the QCD operators $\bar q_{L,R}
\gamma_\mu T^a q_{L,R}$, with  field strengths
\begin{equation}
F_{L,R}^{MN}=F_{L,R}^{MNa} T^a=\partial^M A^N_{L,R} - \partial^N A^M_{L,R}-i[A^M_{L,R},A^N_{L,R}]\,\,\, .
\end{equation}
The gauge fields  enter in the covariant derivative:
$D^MX=\partial^M X-i A_L^M X +i X A_R^M$. Writing  $A_{L,R}$ in
terms of vector $V$ and axial-vector  $A$ fields:
$V^M=\frac{1}{2} (A_L^M+A_R^M)$ and  $A^M=\frac{1}{2}
(A_L^M-A_R^M)$, we obtain the action:
\begin{equation}
S_{eff}=-\frac{1}{k}\int
d^5x\sqrt{-g}\,e^{-\Phi(z)}\,\mbox{Tr}\Big\{|DX|^2+m_5^2X^2+\frac{1}{2g_5^2}\big(F_V^2+F_A^2\big)\Big\}
\label{action1}
\end{equation}
with
\begin{eqnarray}
F_{V}^{MN}&=&\partial^M V^N - \partial^N V^M-i[V^M,V^N]-i[A^M,A^N]\,\,\,,  \non \\
F_{A}^{MN}&=&\partial^M A^N - \partial^N A^M-i[V^M,A^N]-i[A^M,V^N]
\end{eqnarray}
and
$D^MX=\partial^M X-i [V^M ,X] -i \{A^M,X\}$.

The action \eqref{action}-\eqref{action1} is the starting point
of our analysis. Following the AdS/CFT guideline, we assume that the duality relation holds:
\begin{equation}
  \left\langle e^{i\int d^4x\,{\CMcal
  O}(x)f_0(x)}\right\rangle_{QCD}=e^{iS_{eff}}
\end{equation}
where the \emph{lhs} is the QCD generating functional in which
the sources $f_0(x)$ of the $4d$ ${\CMcal O}(x)$ operators are
the boundary ($z\ra0$) limits of the corresponding (dual) $5d$
fields. We then derive the properties of light scalar mesons  on
the basis of the  AdS/CFT duality procedure applied to  the
soft-wall model. The check of duality in this channel is the aim
of the forthcoming Sections.

\section{Spectrum of scalar mesons}
Let us consider the quadratic part of the action
\eqref{action}-\eqref{action1} involving the scalar fields
$S^A(x,z)$:
\begin{equation}
S^{(2)}_{eff}=-\frac{1}{2k}\int
d^5x\sqrt{-g}\,e^{-\Phi(z)}\,\Big(g^{MN}\de_MS^A\de_NS^A+m_5^2S^AS^A\Big)
\,\,\, .
\end{equation}
From this term, it is straightforward to derive  the equation of
motion  for the field $S^A$ (for any flavour index $A$, which is
dropped below):
\begin{equation}
\eta^{MN}\de_M\Big(\frac{R^3}{z^3}\,e^{-\Phi(z)}\de_NS\Big)+3\frac{R^3}{z^5}\,e^{-\Phi(z)}S=0
\end{equation}
or, in the $4d$ Fourier space, defining $S(x,z)=\int \frac {d^4
q}{(2 \pi)^4}\,e^{i q \cdot x} \tilde S(q,z)$ (from now on the
tilde will always denote $4d$ Fourier-transformed fields):
\begin{equation}\label{eqscalar}
\de_z\Big(\frac{R^3}{z^3}\,e^{-\Phi(z)}\de_z\tilde{S}\Big)+3\frac{R^3}{z^5}\,e^{-\Phi(z)}\tilde{S}-q^2\frac{R^3}{z^3}\,e^{-\Phi(z)}\tilde{S}=0\,\,\,.
\end{equation}
 Scalar meson states correspond  to the
normalizable solutions of this equation. The solutions can be
obtained considering the transformation:
\begin{equation}
\tilde S=e^{(\zh^2+ 3 \log {\zh})/2} \, Y
\label{bogolubov}
\end{equation}
with $\zh=cz$ and  the function $Y$ satisfying the one
dimensional Schr\"odinger-like equation
\begin{equation}
-Y'' + V(\zh) Y= \frac{m^2}{c^2}\,Y \,\,\, ; \label{schrodinger}
\end{equation}
the  derivatives act on $\zh$, and the potential is
$V(\zh)=\zh^2+\frac{3}{4 \zh^2}+2$. The normalizable solutions
of eq.\eqref{schrodinger} correspond  to the discrete mass
spectrum \cite{vega}:
\begin{equation}
-q^2_n=m_n^2=c^2(4n+6)
\label{masses}
\end{equation}
with  integer $n$, and eigenfunctions expressed in terms of the
generalized Laguerre polynomials:
\begin{equation}
\tilde S_n(\zh)=\sqrt{\frac{2}{n+1}}\,\zh^3 L_n^1(\zh^2) \,\,\,
. \label{eigen}
\end{equation}
% normalized according to the condition:
%\begin{equation}
%\int_{0}^{\infty}d\hat{z}\frac{e^{-\hat{z}^2}}{\hat{z}^3}|S_n(\hat{z}^2)|^2=1 \,\,\, .
%\end{equation}

The results of this simple calculation can be compared to
current phenomenology. Scalar mesons are organized in linear
Regge trajectories, as a consequence of the choice of the
dilaton field \eqref{dilaton}. The slope of the trajectories is
the same as for vector mesons, the spectral condition of which
is \cite{son2}:
\begin{equation}
m_{\rho_n}^2=c^2(4n+4) \,\,\, .
\label{vectors}
\end{equation}
In the same soft-wall model also scalar glueballs  appear in
Regge trajectories with the same slope, since their masses are
given by \cite{jugeau1}:
\begin{equation}
m_{G_n}^2=c^2(4n+8) \,\,\;;\label{glueballs}
\end{equation}
therefore the parameter $c$ sets the scale of all hadron masses.

Scalar mesons turn out to be heavier than vector mesons. This is
in agreement with experiment if $a_0(980)$ and $f_0(980)$ are
identified as the lightest scalar mesons. The agreement is
quantitative, since eqs.\eqref{masses} and \eqref{vectors} allow
to predict:
$R_{f_0(a_0)}=\frac{m_{f_0(a_0)}^2}{m_{\rho^0}^2}=\frac{3}{2}$,
to be compared to $R_{f_0}^{exp}=1.597 \pm 0.033$ and
$R_{a_0}^{exp}=1.612 \pm 0.004$. Considering the first  radial
exitations, the predictions $R^\prime_{f_0(a_0)}=\frac{5}{4}$
should be compared to the measurements $R_{f_0}^{\prime
exp}=1.06\pm0.04$ and $R_{a_0}^{\prime exp}=1.01 \pm0.04$,
having identified $a_0(1450)$, $f_0(1505)$ and $\rho(1450)$ as
radial excitations; an assignment which however could be
questionable in case of $f_0(1505)$ (identifying $f_0(1370)$
with the first radial excitation, one finds $R_{f_0}^{\prime
exp}=0.9\pm0.2$).

Finally, scalar mesons are lighter than scalar glueballs:
$\frac{m_{G}^2}{m_{f_0}^2}=\frac{4}{3}$ for the lowest-lying
states. Hierarchy among the hadron species  is reduced  for
higher radial states, which become degenerate  when the quantum
number $n$ increases.

\section{Bulk-to-boundary propagator of the scalar field}

According to the AdS/CFT correspondence, the value of the $5d$
field $\wt S(q^2,z)$ at the UV boundary $z=0$, i.e. $\wt
S_0(q^2)$, acts as the source of the corresponding dual $4d$
operator in the QCD functional integral. They are related
through the bulk-to-boundary propagator: $\wt
S(q^2,z)=S(q^2/c^2,\zh^2)\wt S_0(q^2)$. %The bulk-to-boundary
%propagator ${S}(q^2,z)$ of the scalar field allows to relate the
%field  $\tilde{S}(q^2,z)$ to the  source at $z=0$,
%$\tilde{S}_0(q^2)$, of its dual operator:
%$\tilde{S}(q^2,z)=S(\frac{q^2}{c^2}, \zh^2) \tilde{S}_0(q^2)$.
This propagator is obtained solving, for all values of the
four-momenta $q^2$, eq.\eqref{eqscalar} which can be cast in the
form:
\begin{equation}
S''-\frac{1}{\hat{z}}\left(2\hat{z}^2+3\right)S'-\left(\frac{
q^2}{c^2}-\frac{3}{\hat{z}^2}\right)S=0 \label{schrodinger1}
\end{equation}
with the derivatives acting on $\hat{z}$. The general solution
of this equation involves the Tricomi confluent hypergeometric
function $U$ and the Kummer confluent hypergeometric function
${_1F_1}$:
\begin{equation}
S(\frac{q^2}{c^2},\hat{z}^2)=\frac{1}{Rc}\,\G(\frac{
q^2}{4c^2}+\frac{3}{2})\,\hat{z}\,U(\frac{q^2}{4c^2}+\frac{1}{2};
0; \hat{z}^2)+B(\frac{q^2}{c^2})\,\hat{z}^3\,
{_1F_1}(\frac{q^2}{4c^2}+\frac{3}{2}; 2; \hat{z}^2)
\label{btbpropagator}
\end{equation}
with $B(\frac{q^2}{c^2})$ an undetermined function of $q^2/c^2$.
If we impose the boundary condition that the action is finite in
the IR region $z \to +\infty$ (a standard assumption in the
soft-wall model approach) the solution with $B=0$ must be
chosen; we mention below the consequences of relaxing such a
condition, as studied in \cite{jugeau2} for scalar glueballs. In
the UV $z \to 0$ limit, the boundary condition
\begin{equation}
S(\frac{q^2}{c^2},\zh^2)\underset{z\to0}{\to}\frac{z}{R}
\end{equation}
fixes the coefficient of the Tricomi function $U$. With this
expression of the bulk-to-boundary propagator   it is possible
to compute several quantities, namely two- and three-point
correlation functions involving scalar operators.

Before continuing with the analysis, it is worth reminding the
features of the background  field $X_0(z)=\frac{v(z)}{2}$.  It
is solution of the linearized equation of motion:
\begin{equation}
\de_z\Big(\frac{R^3}{z^3}\,e^{-\Phi(z)}\de_zv(z)\Big)+3\frac{R^3}{z^5}\,e^{-\Phi(z)}v(z)=0
\label{cond}
\end{equation}
which explicitely reads:
\begin{equation}\label{cond1}
v(z)=\frac{m_q}{Rc}\,\G(3/2)\,\zh\,
U(1/2;0;\zh^2)+C\,\zh^3\,{_1F_1}(3/2;2;\zh^2) \,\,\, .
\end{equation}
Since both the Tricomi and the Kummer confluent hypergeometric
functions go to unity for  $z \to 0$, the asymptotic UV
behaviour  of \eqref{cond1} is:
\begin{equation}
v(z)\underset{z\to0}{\to}\frac{m_qz}{R}+ \frac{\Sigma z^3}{R}
\end{equation}
with $\Sigma$ related to the constant $C$. Using the AdS/CFT
dictionary, the coefficient of  $z$ enters in the (UV) boundary
condition related to  the quark mass, while the coefficient of
the $z^3$ term is fixed by the (UV) boundary condition related
to the chiral condensate, the two quantities being responsible
of chiral symmetry breaking. However, if one imposes as an  IR
boundary condition that $v(z)$ does not diverge at $z\to
+\infty$, in \eqref{cond1} the solution with $C=0$ must be
chosen, so that the low-$z$ expansion of  $v$ reads:
\begin{equation}\label{vexpand}
v(z)\underset{z\to0}{\to}\frac{m_qz}{R}-\frac{c^2 m_q}{2
R}\,\Big(1-2 \gamma_E-2\ln(cz)-\psi(3/2)\Big)\,z^3 \,\,\,
\end{equation}
with $\psi$ the Euler function. Identifying the coefficient of
the $z^3$ term as the chiral condensate, from eq.\eqref{vexpand}
a proportionality relation can be established between the quark
mass and the quark condensate; this kind of relation is absent
in QCD. This shortcoming, already recognized in the soft-wall
model of AdS/QCD \cite{son2}, does not appear in the hard-wall
model  where the coefficients of $z$ and $z^3$ terms of $v$ are
independent. In principle, it could be avoided by adding
potential terms $U(X)$ to the action \eqref{action}, as
suggested in \cite{son2};  models for  $v(z)$ with the
asymptotic UV and IR behaviour dictated by \eqref{cond} have
also been investigated \cite{lebed2}. In the following, we
ignore this difficulty and use the expression of $v(z)$ in
\eqref{cond1} with $C=0$; the consequences  are important for
the scalar meson couplings to pairs of pseudoscalar states, as
we discuss below \cite{Cherman:2008eh}.

Before concluding this Section,  we report the equations of
motion for the axial and the pion fields. Writing the axial
field $\wt A_\m^a$ in terms of its transverse and longitudinal
components: $\wt A_\m^a=\wt A_{\m\perp}^a + i q_\m \wt\phi^a$,
we have, from the action \eqref{action}-\eqref{action1}:
\begin{eqnarray}
&&\left[\de_z\left(\frac{e^{-\Phi}}{z}\,\de_z \wt A_\m^a\right)-\frac{q^2e^{-\Phi}}{z}\,\wt A_\m^a-\frac{g_5^2\,R^2\,v(z)^2\,e^{-\Phi}}{z^3}\,\wt A_\m^a\right]_\perp=0\label{EOM1}\\
&&\de_z\left(\frac{e^{-\Phi}}{z}\,\de_z\wt\phi^a\right)+\frac{g_5^2\,R^2\,v(z)^2\,e^{-\Phi}}{z^3}\,(\wt\pi^a-\wt\phi^a)=0\label{EOM2}\\
&&q^2\de_z\wt\phi^a+\frac{g_5^2\,R^2\,v(z)^2}{z^2}\,\de_z\wt\pi^a=0\label{EOM3}\;\;\;.
\end{eqnarray}

These equations will be considered below when we compute the scalar meson couplings to pairs of
light pseudoscalar states.

\section{Two-point  correlation function of the scalar operator}
Let us consider in QCD the two-point correlation function:
\begin{equation}
\Pi_{QCD}^{AB}(q^2)= i \int d^4x\,e^{i q \cdot x} \langle
0|T[\CMcal{O}^A_S(x) \CMcal{O}^B_S(0)]|0\rangle \label{twopt}
\end{equation}
with $\CMcal{O}^A_S(x)=\overline{q}(x)T^Aq(x)$. The AdS/CFT
method  relates  this correlation function to the two-point
correlator obtained from the action
\eqref{action}-\eqref{action1}, which  can be written in terms
of the bulk-to-boundary propagator  \eqref{btbpropagator}:
\begin{equation}
\Pi_{AdS}^{AB}(q^2)=\d^{AB}\frac{R^3
c^4}{k}\,S(\frac{q^2}{c^2},\zh^2)\frac{e^{-\Phi(\zh)}}{\zh^3}\,
\de_{\hat z} S(\frac{q^2}{c^2},\zh^2)\Big|_{\zh\to0}
\label{twoptAdS}
\end{equation}
with the result
\begin{eqnarray}
\Pi^{AB}_{AdS}(q^2)&=&\d^{AB}\frac{4c^2R}{k}\,\biggl[\left(\frac{q^2}{4c^2}+\frac{1}{2}\right)\ln(c^2z^2)+
\left(\g_E-\frac{1}{2}\right)+\frac{q^2}{4 c^2}\left(2\g_E-\frac{1}{2}\right)\non \\
&+&\left(\frac{q^2}{4c^2}+\frac{1}{2}\right)\,\psi(\frac{q^2}{4c^2}+\frac{3}{2})\biggr]\biggr|_{\,z=z_{min}}
\,\,\, \label{twoptAdS1}
\end{eqnarray}
(omitting a $\CMcal{O}(\frac{1}{z^2})$ contact term). The AdS
expression of the correlation function (plotted in
fig.\ref{fig:twopt}) shows the presence of a discrete set of
poles, corresponding to the poles of the Euler function $\psi$,
with masses  given by the spectral relation \eqref{masses}
%%%%%%%%%%%%%%%%%%%%%%%%%%%%%%%%%%%%%%%%%%
\begin{figure}[t]
\begin{center}
\vspace*{0.2cm}
\includegraphics[width=0.50\textwidth] {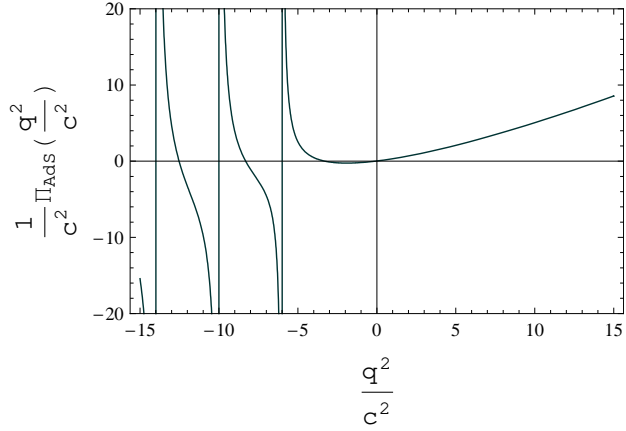}
\end{center}
\caption{\baselineskip=14pt  Two-point correlation function
$\frac{1}{c^2} \Pi_{AdS}(q^2/c^2)$ in eq.\eqref{twoptAdS1}. The
renormalization scale is fixed to $\nu=1$ GeV.}
\label{fig:twopt}
\end{figure}
and residues
\begin{equation}
F_n^2=\frac{R}{k}\,16\,c^4(n+1) \,\,\, . \label{residues}
\end{equation}
The factor $\frac{R}{k}$ can be fixed by matching the AdS
expression  \eqref{twoptAdS1} in the $q^2\to +\infty$ (i.e. in
the short-distance) limit, expanded in powers of $1/q^2$,  with
the QCD result. For $\Pi_{AdS}$, identifying  $z_{min}$ with the
renormalization scale $\frac{1}{\n}$, we
get:
\begin{eqnarray}
\Pi^{AB}_{AdS}(q^2)&=&\d^{AB}\frac{R}{k}
\left[q^2\ln(\frac{q^2}{\n^2})+q^2\left(2\g_E-\ln4-\frac{1}{2}\right)+2c^2\left(\ln(\frac{q^2}{\n^2})-\ln4+2\g_E+1\right)\right.\non \\
&+&\left.\frac{2}{3}\frac{c^4}{q^2}+\frac{4}{3}\frac{c^6}{q^4}+O(1/q^6)\right]
\label{twoptAdS2}
\end{eqnarray}
while the QCD result, for $N_c=3$ and in terms of quark and
gluon condensates, is \cite{reinders}:
\begin{eqnarray}
\Pi^{AB}_{QCD}(q^2)&=&\frac{\delta^{AB}}{2} \left[
\frac{3}{8\pi^2}\left(1+\frac{11 \alpha_s}{3 \pi}\right)
q^2\ln(\frac{q^2}{\n^2})+\frac{3}{q^2}\,\langle
m_q\overline{q}q\rangle+\frac{1}{8 q^2}\,\langle
\frac{\a_s}{\pi}
G^2\rangle\right.\non\\
&+&\left.\frac{m_qg_s}{2q^4}\,\langle (\overline{q}\s_{\m\n}\l^a
q)G^a_{\m\n}\rangle+\frac{\pi\a_s}{q^4}\,\langle
(\overline{q}\s_{\m\n}\l^a
q)^2\rangle\right. \non \\
 &+&\left.\frac{2\pi\a_s}{3q^4}\,\langle(\overline{q}\g_{\m}\l^aq)\sum_{q=u,d}\overline{q}\g_{\m}\l^aq\rangle
+ O(1/q^6)\ \right] \,\,\, \label{twoptQCD2}
\end{eqnarray}
where possible terms related, e.g., to instanton contributions
have not been considered.
Matching the perturbative term fixes the condition:
\begin{equation}
\frac{R}{k}=\frac{N_c}{16\pi^2} \,\,\ .
\label{match}
\end{equation}
In the same way, by the two-point correlation function of the
vector current, the matching condition  fixes the value of
$g_5^2$: $g_5^2=3/4$.

The residues of the two-point correlation function, related to
the scalar meson decay constants, are now determined:
\begin{equation}
F_n^2=\frac{N_c}{\pi^2}\,c^4(n+1) \,\,\,
\label{residues1}
\end{equation}
for all radial states labeled by $n$.

It is interesting to compare \eqref{residues1} to QCD
calculations. For $a_0(980)$, the following result has been
obtained for the current-vacuum matrix elements defining the
decay constants:
\begin{equation}
F_{a_0}=\langle 0| \CMcal{O}^3_S|\,a_0(980)^0\rangle = (0.21 \pm
0.05) \,\, {\rm GeV}^2 \,\,\, \mbox{\cite{gokalp}}\,\,\,.
\end{equation}
The AdS prediction is: $F_{a_0}=\frac{\sqrt
3}{\pi}c^2=0.08$~GeV$^2$, having fixed $c$ from the $\rho^0$
mass: $c=\frac{m_\rho}{2}$. For the $f_0(980)$ a similar result
has been obtained for the matrix element of the $s\bar s$
operator: $\langle 0|\,\bar s s\,| f_0(980)\rangle = (0.18 \pm
0.015)$~GeV$^2$ \cite{defazio}. The AdS result is not far from
QCD determinations. For the first radial excitation we have:
$F_{a_0^\prime}=0.12$ GeV$^2$, while for large values of $n$ the
ratio $\frac{F_n^2}{m_n^2}$ becomes independent of the radial
quantum number.

AdS/QCD duality can be checked for the various terms in the
$\frac{1}{q^2}$ power expansion, comparing eqs.\eqref{twoptAdS2}
and \eqref{twoptQCD2}. For $m_q=0$, the four dimensional gluon
condensate can be computed:
\begin{equation}
\langle\frac{\a_s}{\pi}G^2\rangle=\frac{2}{\pi^2}c^4\simeq0.004\,\,
{\rm GeV}^4
\end{equation}
which is smaller than the commonly used value $
\langle\frac{\a_s}{\pi}G^2\rangle\simeq0.012\,$ GeV$^4$, the
estimated uncertainty of which is about $30\%$
\cite{khodjamirian}.

Considering $O(1/q^4)$ terms, in QCD one can use the
factorization approximation:
\begin{equation}\label{factorization approx}
\begin{array}{lll}
\langle(\overline{q}\s_{\m\n}\l^aq)^2\rangle&\simeq&-\frac{16}{3}\langle\overline{q}q\rangle^2\,\,\,,\\
\\
\langle(\overline{q}\g_{\m}\l^aq)^2\rangle&\simeq&-\frac{16}{9}\langle\overline{q}q\rangle^2
\end{array}
\end{equation}
for the dimension 6 operators. Within  such an approximation,
the AdS and QCD expressions do not match, since the $O(1/q^4)$
term in \eqref{twoptAdS2} is positive, while it is negative in
\eqref{twoptQCD2}.

The last remark is that in the AdS expression \eqref{twoptAdS2}
there is a contribution interpreted in terms of a dimension two
condensate, while  an analogous term is absent in the QCD
expansion \eqref{twoptQCD2}. In this respect, the two-point
correlation function of the scalar operators presents the same
phenomenon occurring in the two-point correlation function of
vector mesons \cite{andreev1,Beijing} and of scalar glueballs
\cite{forkel,jugeau2}.  Although in QCD there is no local
gauge-invariant operator of dimension two, the possible
relevance of a dimension two condensate in the form of an
effective gluon mass term is the subject of  discussions
\cite{zakharov}, so that the AdS result could be interpreted as
an argument supporting  the existence of this condensate.
However,  the AdS/CFT method dictates duality between bulk
fields and gauge-invariant operators in the boundary theory.
Another possible  way to explain the presence of  this
contribution is that, although the quadratic dependence of the
dilaton field in the IR  is required  to provide linear
confinement, at smaller values of $z$ the functional dependence
of $\Phi(z)$ is less constrained, so that in other versions of
the background field such a term could be removed: this deserves
an explicit check. A different possibility, put forward in
\cite{jugeau2}, is that the subleading (for $z \to 0$) solution
in the bulk-to-boundary scalar field propagator plays a role, so
that its coefficient can be tuned to cancel the dimension two
contribution. In such a scenario, in which the AdS dual theory
needs to be regularized in the IR, the subleading solution
modifies some terms in the power expansion of the two-point
correlation function, leaving the perturbative term unaffected.

\section{Interaction of scalar mesons  with a pair of pseudoscalar mesons}

In  the action \eqref{action} the interaction terms involving
one scalar $S$ and two light pseudoscalar fields $P$  only
appear in the covariant derivative Tr$\Big\{|DX|^2\Big\}$.
Using  the equations of motion and writing the axial-vector
bulk field in terms of the transverse and longitudinal
components: $A_M=A_{\perp\,M}+\de_M\phi$, we have:
\begin{equation}\label{spipi}
S^{(SPP)}_{eff}=-\frac{4}{k}\int
d^5x\sqrt{-g}\,e^{-\Phi(z)}g^{MN}
v(z)\,\mbox{Tr}\Big\{S(\de_M\pi-\de_M\phi)(\de_N\pi-\de_N\phi)\Big\}
\end{equation}
i.e.,
\begin{eqnarray}\label{spipi1}
S^{(SPP)}_{eff}&=&-\frac{4}{k}\int
d^5x\sqrt{-g}\,e^{-\Phi(z)}g^{MN}v(z)\,S_1(\de_M\psi^a)(\de_N\psi^b)\frac{1}{\sqrt{2n_F}}\,\mbox{Tr}[T^aT^b]\non\\
&&-\frac{4}{k}\int
d^5x\sqrt{-g}\,e^{-\Phi(z)}g^{MN}v(z)\,S_8^a(\de_M\psi^b)(\de_N\psi^c)\,\mbox{Tr}[T^aT^bT^c]
\,\,\,
\end{eqnarray}
where  $\psi^a=\phi^a-\pi^a$.
For $n_F=2$,  Tr$[T^aT^bT^c]=\frac{i}{4}\ve^{abc}$ and the octet
(triplet) part vanishes,  while for $n_F=3$  we have:
\begin{eqnarray}
S^{(SPP)}_{eff}&=&-\frac{R^3}{k}\frac{2}{\sqrt{6}}\int
d^5x\,\frac{1}{z^3}\,e^{-\Phi(z)}v(z)\,S_1\,\eta^{MN}(\de_M\psi^a)(\de_N\psi^a) \non \\
&&-\frac{R^3}{k}d^{abc}\int
d^5x\,\frac{1}{z^3}\,e^{-\Phi(z)}v(z)\,S_8^a\,\eta^{MN}(\de_M\psi^b)(\de_N\psi^c)
\,\,\, .
\end{eqnarray}
In the Fourier space,  this term involves the bulk-to-boundary
propagator $S(\frac{q^2}{c^2},c^2z^2)$ of the scalar field,
together with the sources $\tilde S_{1(8)_0}$.  The longitudinal
part of the axial-vector field can be related to its source
through the equation:
\begin{equation}
{\wt \phi}^a(q,z)=\frac{1}{q^2}\,{\CMcal
A}_\| (\frac{q^2}{c^2},c^2z^2)(-iq^{\m}\tilde{A}^a_{{\|_{\,0}}\,\m}(q))
\,\,\, ,
\end{equation}
while for the combination
$\wt \psi^a= \wt \phi^a- \wt \pi^a$ the equation involves the propagator $\Psi$:
\begin{equation}
{\wt \psi}^a(q,z)=\frac{1}{q^2}\,\Psi(\frac{q^2}{c^2},c^2z^2)(-iq^{\m}\tilde{A}^a_{{\|_{\,0}}\,\m}(q)) \,\,\,.
\end{equation}
The contribution of only the  pseudo-Goldstone bosons is selected by the condition:
\begin{equation}
  {\wt \psi_P}^a(q,z)=\frac{1}{q^2}\,\Psi(0,c^2z^2)(-iq^{\m}\tilde{A}^a_{{\|_{\,0}}\,\m}(q))
\,\,\,.
\end{equation}
From eq. \eqref{EOM3} the condition  $\de_z \wt \pi^a=0$ holds
at $q^2=0$ and the equation for $\Psi(0,c^2z^2)$:
\begin{equation}\label{EOMpi-phi}
  \de_z\left[\frac{e^{-\Phi}}{z}\,\de_z\Psi(0,c^2z^2)\right]-\frac{g_5^2\,R^2\,v(z)^2\,e^{-\Phi}}{z^3}\,\Psi(0,c^2z^2)=0\;\; \,\,\,
\end{equation}
 coincides with the  equation holding for  ${\CMcal A}(0,c^2z^2)$ which appears in the relation
 $\wt A_{\perp\m}^a(0,z) =  {\CMcal A}(0,c^2z^2) \wt A^a_{ \perp 0 \m}(0)$.
 We can then identify
$\Psi(0,c^2z^2)= {\CMcal A}(0,c^2z^2)$ \cite{Grigoryan:2007wn}, so that:
\begin{equation}
  {\wt \psi_P}^a(q,z)=\frac{1}{q^2}\,{\CMcal A}(0,c^2z^2)(-iq^{\m}\tilde{A}^a_{{\|_{\,0}}\,\m}(q))
\,\,\,.
\end{equation}
In this way,  the $S_{eff}^{(SPP)}$ term in \eqref{spipi1},
considering only the octet contribution, reads:
\begin{eqnarray}\label{Spipi}
iS^{(SPP)}_{eff}&=&-\frac{i}{k}\,d^{abc}\int\frac{d^4q_1d^4q_2d^4q_3}{(2\pi)^{12}}\,(2\pi)^4\d^4(q_1+q_2+q_3)\times\non\\
&&\int_{0}^{\infty}
dz\,\frac{R^3}{z^3}\,e^{-\Phi(z)}v(z)\,S(\frac{q^2_1}{c^2},c^2z^2)\,\tilde{S}^a_{8_0}(q_1)
\Big[\big(\de_z\A(0,c^2z^2)\big)^2-q_2\cdot q_3\,\A(0,c^2z^2)^2\Big]\times\non\\
&&\Big(-\frac{i}{q^2_2}\,q^{\m}_2\tilde{A}^b_{{\|_{\,0}}\,\m}(q_2)\Big)\Big(-\frac{i}{q^2_3}\,q^{\n}_3\tilde{A}^c_{{\|_{\,0}}\,\n}(q_3)\Big)\,\,\,.
\end{eqnarray}
This interaction term allows to compute the scalar couplings to
pseudoscalar states. Indeed, on the basis of the AdS/CFT
correspondence, the QCD three-point correlation function
involving two pseudoscalar and one scalar operator:
\begin{equation}\label{3ptQCD}
\Pi_{QCD\a \b}^{abc}(p_1,p_2)=i^2 \int d^4x_1 d^4x_2\,e^{i
p_1\cdot x_1} e^{i p_2\cdot x_2}
\langle0|T[\CMcal{O}^b_{5_{\a}}(x_1)\CMcal{O}^a_S(0)\CMcal{O}^c_{5_{\b}}(x_2)]|0\rangle
\,\,\,
\end{equation}
can be obtained by functional derivation of \eqref{Spipi} with
respect to the source fields $\tilde{A}_{\parallel_{\,0}}(p_1)$,
$\tilde{A}_{\parallel_{\,0}}(p_2)$ and $\tilde{S}^a_{8_0}(q)$, with the result:
\begin{eqnarray}\label{dualityrelation}
&&\Pi_{AdS\a \b}^{abc}(p_1,p_2)=\non \\
&&\frac{p_{1\a}p_{2\b}}{p_1^2p_2^2}\frac{2
R^3}{k}d^{abc}\int_0^{\infty}dz\frac{1}{z^3}\,e^{-\Phi}v(z)S(\frac{q^2}{c^2},c^2z^2)\left[\left(\de_z\A(0,c^2z^2)\right)^2-\frac{q^2}{2}\A(0,c^2z^2)^2\right]
\end{eqnarray}
with $q=-(p_1+p_2)$. The AdS expression of the strong $SPP$
couplings follows writing the bulk-to-boundary propagator $S$ in
terms of the scalar mass poles, of the residues and of the
normalizable eigenfunction $\tilde S_n(\zh^2)$ in \eqref{eigen}.
Using the integral representation of the Tricomi  function
\cite{Beijing,Erdelyi}:
\begin{equation}
  U(a,b,x)=\fr{1}{\G(a)}\int_0^1dy\,\fr{y^{a-1}}{(1-y)^b}\,\exp\left[-\fr{y}{1-y}\,x\right]\,\,\,,
\end{equation}
one derives the generating
function of the Laguerre polynomials \cite{radyushkin}:
\begin{equation}
  \fr{1}{(1-y)^2}\,\exp\left[-\fr{y}{1-y}\,x\right]=\sum_{n=0}^{\infty}L_n^1(x)\,y^n
\end{equation}
so that:
\begin{equation}
S(\frac{q^2}{c^2},c^2z^2)=\frac{1}{Rc}\,\sqrt{\frac{8}{N_c}}\;\pi\sum_{n=0}^{\infty}\frac{F_n
\tilde S_n(c^2z^2)}{q^2+m_n^2+i\ve} \,\,\,.
\end{equation}
Moreover, defining the scalar form factor $F_{P}$:
\begin{equation}
\langle P^{d}|\CMcal{O}_S^a|P^e\rangle=F^{dae}_{P}\big(q^2\big) \,\,\, ,
\end{equation}
we have:
\begin{equation}
\Pi^{abc}_{QCD \a \b}(p_1,p_2)=
-\frac{p_{1\a}p_{2\b}}{p_1^2p_2^2}f_{\pi}^2\,F_{P}^{abc}\big(q^2\big)
\,\,\,.
\end{equation}
The AdS expressions of  the scalar form factor and of the
$g_{S_n PP}$ couplings follow:
\begin{eqnarray}
F_{P}^{abc}(q^2)&=&-d^{\,abc}\frac{1}{k}\frac{2}{f_{\pi}^2}\int_0^{\infty}dz
\frac{R^3}{z^3}\,e^{-\Phi}v(z)S(\frac{q^2}{c^2},c^2z^2)
\left[\left(\de_z\A(0,c^2z^2)\right)^2-\frac{q^2}{2}\A(0,c^2z^2)^2\right]
\non \\
&=&-d^{\,abc}\sum_{n=0}^{\infty}\frac{F_ng_{S_nPP}}{q^2+m_n^2}
\end{eqnarray}
with
\begin{equation}\label{gspipicoup}
g_{S_n
PP}=\frac{1}{k}\frac{2}{f_{\pi}^2}\int_0^{\infty}dz\frac{R^3}{z^3}\,e^{-\Phi}v(z)\frac{1}{Rc}\sqrt{\frac{8}{N_c}}\;\pi
\tilde
S_n(c^2z^2)\Big[\big(\de_z\A(0,c^2z^2)\big)^2+\frac{m_{S_n}^2}{2}\A(0,c^2z^2)^2\Big]
\,\,\, .
\end{equation}

To compute $g_{S_nPP}$ from \eqref{gspipicoup}, one needs
 $\A(0,c^2z^2)$, which can be obtained solving  \eqref{EOM1}. However, since $v(z)$ is small
(it depends on $m_q/R$), one can
neglect terms  proportional to $v^2$ and identify
$\A(0,\zh^2)$ with $\A^{(0)}(0,\zh^2)$ solution of:
\begin{equation}
\de_{\zh}\left(\frac{e^{-\zh^2}}{\zh}\,\de_{\zh}\A^{(0)}(0,\zh^2)\right)=0\;\;
\end{equation}
with $\A^{(0)}(0,\zh^2)\underset{z\to0}{\to}1$. The regular
solution is  $\A^{(0)}(0, \zh^2)=1$.

The expression of $g_{S_0PP}$ for the lowest radial number
$n=0$, since $\tilde S_0(\zh^2)=\sqrt{2} \zh^3$, is:
\begin{equation}
g_{S_0PP}=\frac{\sqrt{N_c}}{4\pi}\frac{m_{S_0}^2}{f_{\pi}^2} Rc
\int_{0}^{\infty}d\hat{z}\,e^{-\hat{z}^2} v(\hat{z})  \,\,\, .
\end{equation}
The coupling depends linearly on the  field $v$. The numerical
result is   small, of the order of $10$ MeV depending on the
quark mass used as an input. On the other hand, phenomenological
determinations of the $SPP$ couplings indicate sizeable values,
showing that  the scalar states are characterized by their large
couplings to light pseudoscalar mesons. For example, the
experimental value of $g_{a_{0}\eta \pi}$ is: $g_{a_{0}\eta
\pi}=12\pm6$ GeV, while for $f_0$ the result of a QCD estimate
is: $g_{f_0 K^+ K^-}\simeq 6-8$ GeV \cite{defazio3}. The origin
of the  small value for the $SPP$ couplings in the AdS/QCD
soft-wall model can be traced to the expression of $v$ which is
determined by the light quark mass (larger results would be
obtained, e.g.,  using $v$ computed in the hard-wall approach).
The drawback confirms the difficulty of the  soft-wall model in
correctly describing  chiral symmetry breaking; it could be
probably avoided including potential terms in the effective
action \eqref{action}, a possibility which deserves a dedicated
study.

\section{Conclusions}
We have studied the scalar sector in the $5d$ AdS soft-wall
model proposed as a QCD dual, finding  that the masses and decay
constants of scalar mesons are close to experiment and QCD
determinations. The two-point correlation function of the scalar
operator has a power expansion similar to QCD, with violations
in the dimension six condensates computed assuming
factorization. A dimension two condensate  term, absent in QCD,
appears in the power expansion of the AdS expression,
analogously to the two-point correlators of vector meson and
scalar glueball operators. The strong couplings of scalar states
to pairs of light pseudoscalar mesons are smaller
than in phenomenological determinations, as a consequence of the
difficulty   of correctly describing chiral symmetry breaking
within this model.  This difficulty could be avoided including
additional  potential terms  in  the effective Lagrangian
defining the  model.

\vspace{1cm}

\noi {\bf Acknowledgments.}\\
\noi This paper is dedicated to the memory of Beppe Nardulli.

\noi We are grateful to M.~Pellicoro for discussions. This work
was supported in part by the EU Contract No.
MRTN-CT-2006-035482, "FLAVIAnet".

\end{document}